\documentclass[epj]{webofc}
\usepackage[utf8]{inputenc}
\usepackage[varg]{txfonts}   
\usepackage{booktabs}
\usepackage{xcolor}
\definecolor{darkred}{rgb}{0.4,0.0,0.0}
\definecolor{darkgreen}{rgb}{0.0,0.4,0.0}
\definecolor{darkblue}{rgb}{0.0,0.0,0.4}
\usepackage{subfigure}%
\wocname{EPJ Web of Conferences}
\woctitle{Lattice2017}
\begin{document}
%
\selectlanguage{english}
\title{%
MILC Code Performance on High End CPU and GPU \\
Supercomputer Clusters
}
\author{%
\firstname{Carleton} \lastname{DeTar}\inst{1} \and
\firstname{Steven} \lastname{Gottlieb}\inst{2}\fnsep\thanks{Acknowledges financial support by Intel Parallel Computing Center at Indiana University.} \and
\firstname{Ruizi}  \lastname{Li}\inst{2}\fnsep\thanks{Speaker, \email{ruizli@umail.iu.edu}}\and \firstname{Doug}  \lastname{Toussaint}\inst{3}}

\institute{%
Department of Physics \& Astronomy, University of Utah, Salt Lake City, UT 84112, U.S.A.
\and
Department of Physics, Indiana University, Bloomington, IN 47405, U.S.A.
\and
Department of Physics, University of Arizona, Tucson, AZ 85721, U.S.A.
}
\abstract{%
With recent developments in parallel supercomputing architecture, 
many core, multi-core, and GPU processors
are now commonplace, resulting in more levels of parallelism, memory
hierarchy, and programming complexity.  It has been necessary to adapt
the MILC code to these new processors starting with NVIDIA
GPUs, and more recently, the Intel Xeon Phi processors.  We report on 
our efforts to port and optimize our code for the Intel Knights Landing
architecture.  We consider performance of the MILC code with MPI and OpenMP, 
and optimizations with QOPQDP and QPhiX.  For the latter approach, we 
concentrate on the staggered conjugate gradient and gauge force.  We also
consider performance on recent NVIDIA GPUs using the QUDA library.
}
\maketitle
\section{Introduction}\label{intro}


The MILC code has been in production and freely available for over 20 years,
with continual improvements to match our evolving
physics goals and changing hardware.
Currently a code of approximately 180,000 lines, it is
used by several collaborations worldwide.
Originally, there was a single level of parallelization based on message
passing.  With the advent of MPI, that became the main message passing 
library used, though there are some others as mentioned below.  OpenMP
parallelization was briefly tried around 2000, but is now more fully
developed.  The code has made increasing use of a library of specialized
data-parallel linear algebra and I/O routines developed over the past
several years with support from the DOE's Scientific Discovery through Advanced Computing (SciDAC) Program.  
In addition, we make use of the QUDA library for computers with NVIDIA GPUs.
We have been porting the code to the Intel Xeon Phi many-integrated-core (MIC) architecture named Knights Landing (KNL).
This architecture contains 512-bit SIMD vector processors that can 
do floating point arithmetic on 16 single precision or 8 double 
precision numbers at a time.
To exploit the power of these vector units, we have been generalizing
the QPhiX library \cite{QPhiX} for Wilson/Clover quarks developed
by Jefferson Lab and Intel to support staggered quarks.
Currently, the staggered QPhiX library contains code for multi-shift
and single mass conjugate gradient solvers.
In addition, a Symanzik one-loop improved gauge force is available for
gauge field generations.  Routines to support smearing for HISQ quarks are
under development.
In addition, we are improving the OpenMP parallelization in the MILC code. \\
\\
In the rest of this paper, we will briefly review the development of 
the MILC code, describe the three main libraries that the code can call to
enhance performance, and then present a number of benchmarks.  Finally, we
present some conclusions.

\section{Development of the MILC code}\label{sec-1}

The MILC code was originally developed to allow single processor or
multiprocessor (with message passing) running in a way that is transparent
to the developer.  Selecting a MILC communication file at compile time
was the only difference between serial and parallel running.  All of
the application code was otherwise identical.  Further, there were several
message passing libraries at the time, so MILC had one for each message
passing library, plus {\tt com\_vanilla.c} that did not need to pass messages
but implemented the MILC calls like {\tt start\_gather}, {\tt wait\_gather}, and 
{\tt cleanup\_gather}.  This isolated our users and developers from having to
understand multiple message passing systems.  Eventually, MPI won out.
Later, OpenMP was supported to make use of 
symmetric multiprocessing (SMP) machines with the shared memory \cite{MILCwMPI}; 
however, this required examining individual {\tt FORALLSITES} loops and new
{\tt FORALLSITES\_OMP}.  Only with the advent of the Intel MIC architecture
this did again become an area of focus.\\
\\
Support for GPUs is based on the QUDA library \cite{QUDA}, originally developed at
Boston University (BU).  The extension to staggered quarks began when one of us
was on sabbatical at the National Center for Supercomputing Applications
(NCSA) at the University of Illinois.  The QUDA community has grown over 
the years and benefits greatly from the leadership of NVIDIA
staff members Kate Clark, one of the original BU developers, and Mathias Wagner,
a former MILC postdoc.

\section{Libraries: SciDAC, QUDA, staggered QPhiX}\label{sec-1}

The basic MILC code can call three packages to improve performance.
USQCD's SciDAC-funded software package is used by the MILC code for 
runs on various CPU architectures.  The software packages include
QLA for basic linear algebra, QDP for data parallel routines,
and QOPQDP, the high-level optimized library that supports functions
such as CG solvers, gauge force, etc.
The SciDAC package can use SSE2 instructions, but not newer SIMD
instructions sets such as AVX512.\\
\\
To make use of GPUs on machines with NVIDIA hardware, the MILC code calls
the QUDA library which has all of the routines needed for creating new
gauge fields.  Of course, that includes the solvers required to make quark
propagators in analysis codes.  QUDA supports mixed-precision calculation
as well as more than one method of gauge-field compression to improve
performance.
The version benchmarked here is 0.8.0. \\
\\
We have been developing the staggered QPhiX library 
as part of our effort to port the MILC code to the MIC architecture. 
QPhiX targets the first and second generations of this architecture. 
It also supports other instruction sets, e.g., AVX2, SSE. 
This library 
supports SMID vectorization via intrinsics, and OpenMP threading.  As
each MPI task can call the library, the resulting code supports all
three levels of parallelization available on the Intel Xeon Phi, or the
new Skylake chips.
A detailed description of the QPhiX library framework can be found at \cite{QPhiX}. 
The library uses a structure-of-array (SOA) data structure for improved cache reuse. 
The data layout is the same as in the Grid library \cite{Grid}. 

\section{Benchmarks}\label{sec-1}

\begin{figure}[tp]
   \centering
   {\includegraphics[width=0.7\textwidth,height=0.38\textheight,clip]{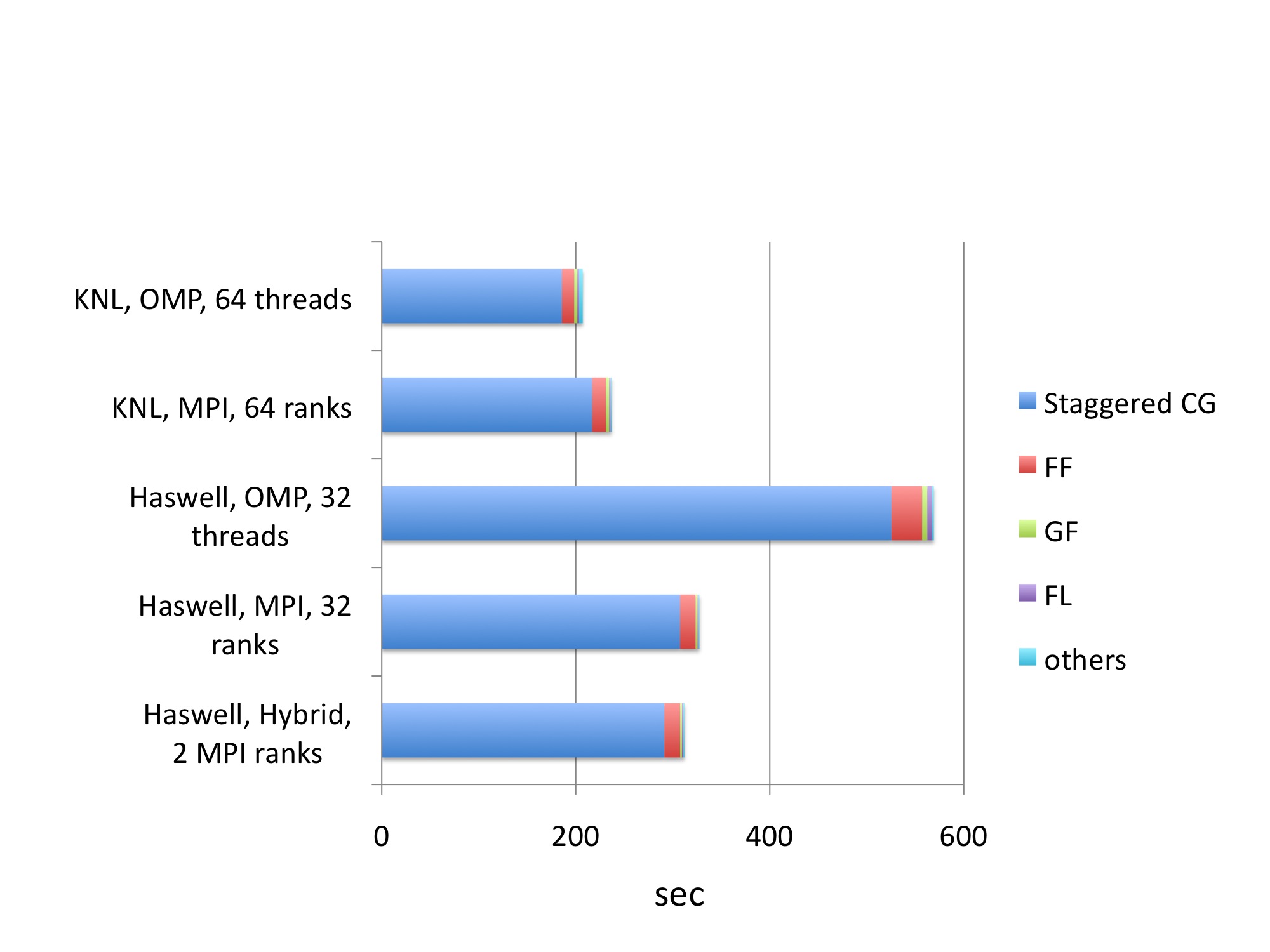}}\hfill
   \caption{Runtime breakdown (in seconds) for the current version of the MILC code that also includes improved OpenMP parallelizations, 
on a single KNL 7250 or Haswell dual-socket 16-core Intel Xeon E5-2698 v3 node, 
   with MPI, OpenMP, and hybrid MPI/OpenMP parallelism. 
   The bars in red, green, and purple denote the HISQ fermion force (FF), 
Symanzik gauge force (GF), and HISQ link smearing (FL) time, respectively. 
   The lattice size is $16^4$. 
   No hyper-threading was enabled, 
   and the run on Haswell with hybrid parallelism has 16 threads per rank.}
   \label{Fig.1}
\end{figure}

The benchmarks shown here use the {\tt su3\_rhmd\_hisq} application in the MILC code.
This code implements gauge-field evolution using the RHMD algorithm for
HISQ quarks.
All results present here are for double precision. 
As expected, single-precision code performance is about twice as fast.\\
\\
In Figure~\ref{Fig.1}, we present a bar chart showing how time is spent in
various parts of the code using various hardware and software combinations.
Starting from a reasonably equilibrated $16^4$ configuration,
we see that about $90\%$ of the time is spent in the CG routine, which is
shown in blue. 
The HISQ fermion force, gauge force, and HISQ-link smearing are shown using
other colors explained in the legend and caption. 
The top two bars are for runs on a KNL node with a single chip and 
the other three are for a dual socket Intel Haswell node.
The slowest performance is from a pure OpenMP run on the dual-socket Haswell.
However, a hybrid MPI/OpenMP run with one MPI task per socket beats
a pure MPI run with 32 MPI tasks.
On the KNL node, pure OpenMP with 64 threads is slightly faster than MPI
with 64 ranks.

\subsection{Staggered multi-shift CG}\label{sec-2}

As mentioned, the staggered multi-shift CG is the most time-consuming part 
of the code in production runs. 
We carried out a set of weak-scaling benchmarks of this routine. 
All of our KNL benchmarks were collected on three clusters: Stampede~2 \cite{Stampede} at the
Texas Advance Computing Center (TACC), Theta \cite{Theta} at the Argonne Leadership
Computing Center (ALCF), and Cori~II \cite{Cori} at the National Energy Research 
Supercomputing Center (NERSC). 
Each of these clusters has KNL nodes with either 64 or 68 cores per node, 
Theta and Cori~II use the Cray Aries network, whereas Stampede~2
uses the Intel Omni-path network. 
The maximum number of nodes used was 2048, on Cori~II, 
covering around half of the entire cluster at the time of run. 
Results for the baseline code below refer to the MILC code with MPI only, unless noted otherwise. \\

\begin{figure}[tp]
   \centering
   \subfigure[Baseline code on three KNL clusters, 
   shows peak performance 
   at lattice size between $L = 24$ and $36$. 
             ]%
             {\includegraphics[width=0.30\textwidth,clip]{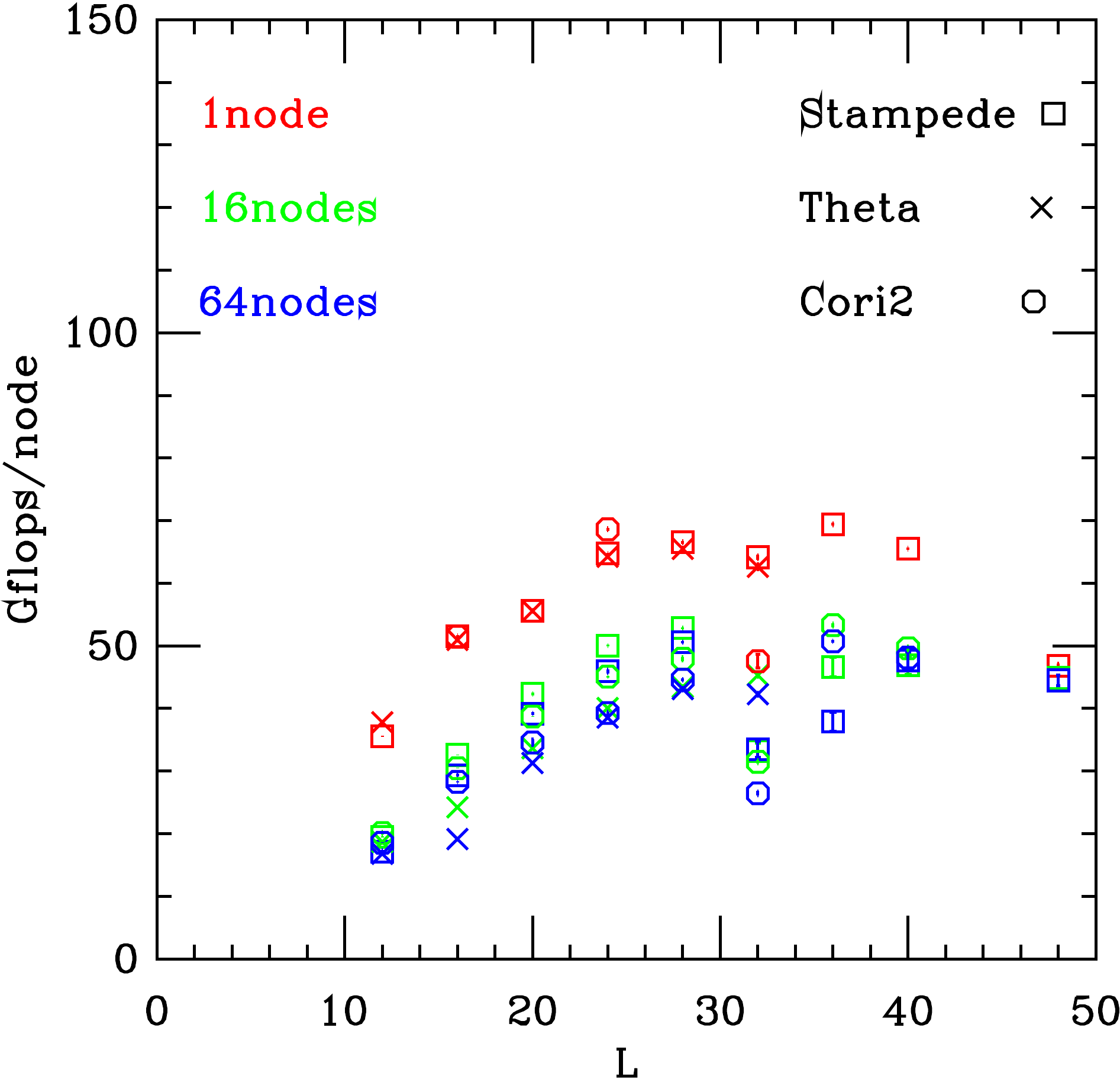}} \hspace{1cm} 
   \subfigure[Code with QPhiX on two KNL clusters. 
   One node runs are with OpenMP, 
   and multe-node runs are with up to 16 ranks per node. 
   ]%
             {\includegraphics[width=0.30\textwidth,clip]{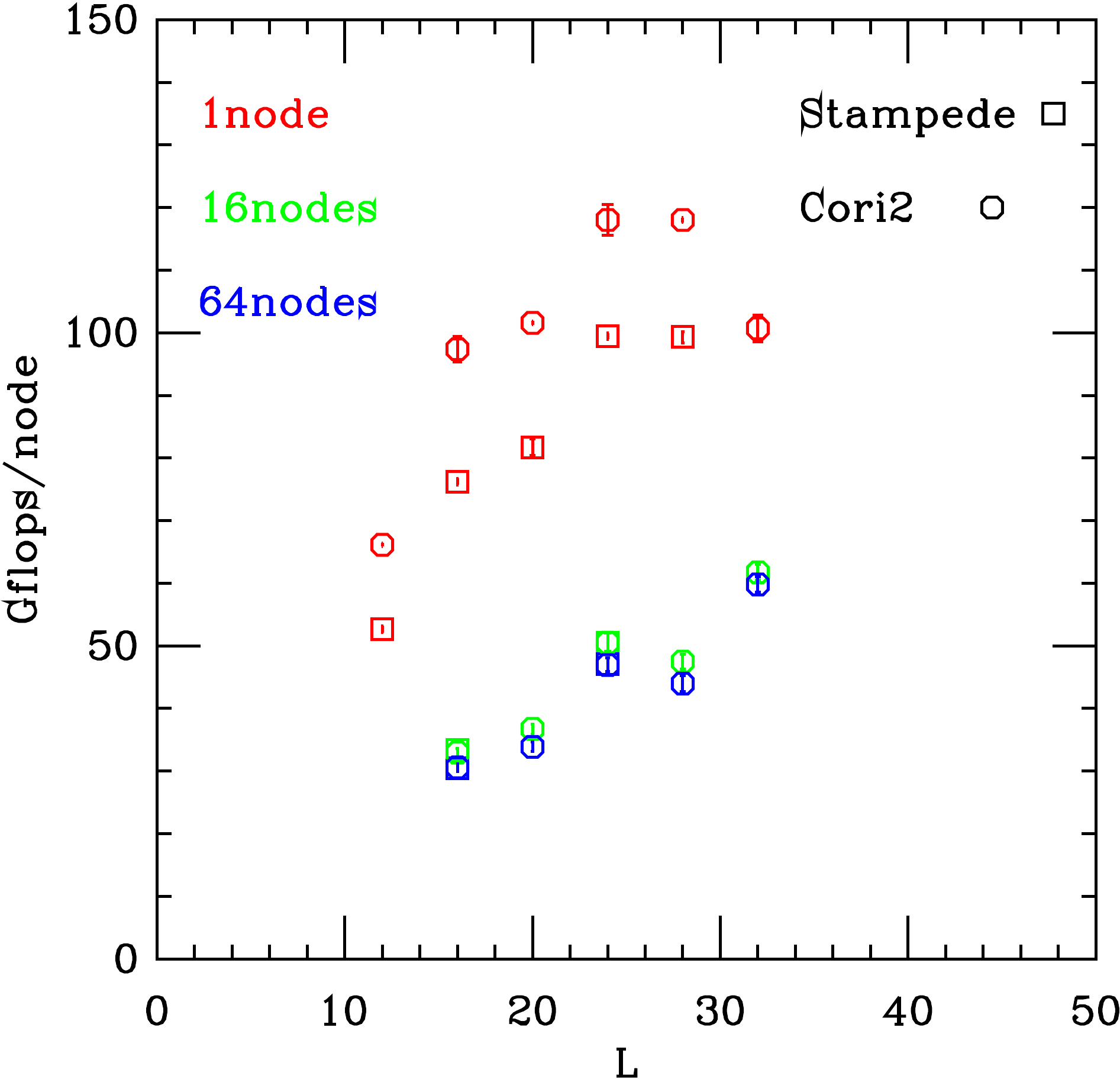}} \\
   \caption{Staggered multi-shift CG weak-scaling performance on KNL clusters up to 64 nodes, with baseline code (left) and QPhiX (right). 
   The horizontal axis encodes the lattice size per node $L^4$, and the vertical axis is the flop rate per node. 
   The performance improvement with QPhiX can be over $50\%$ on one node, but much less on multiple nodes. 
   }
   \label{Fig.2}
\end{figure}

\begin{figure}[tp]
   \centering
   \subfigure[Baseline code performance on multiple nodes is higher with more MPI ranks per node. 
             ]%
             {\includegraphics[width=0.32\textwidth,clip]{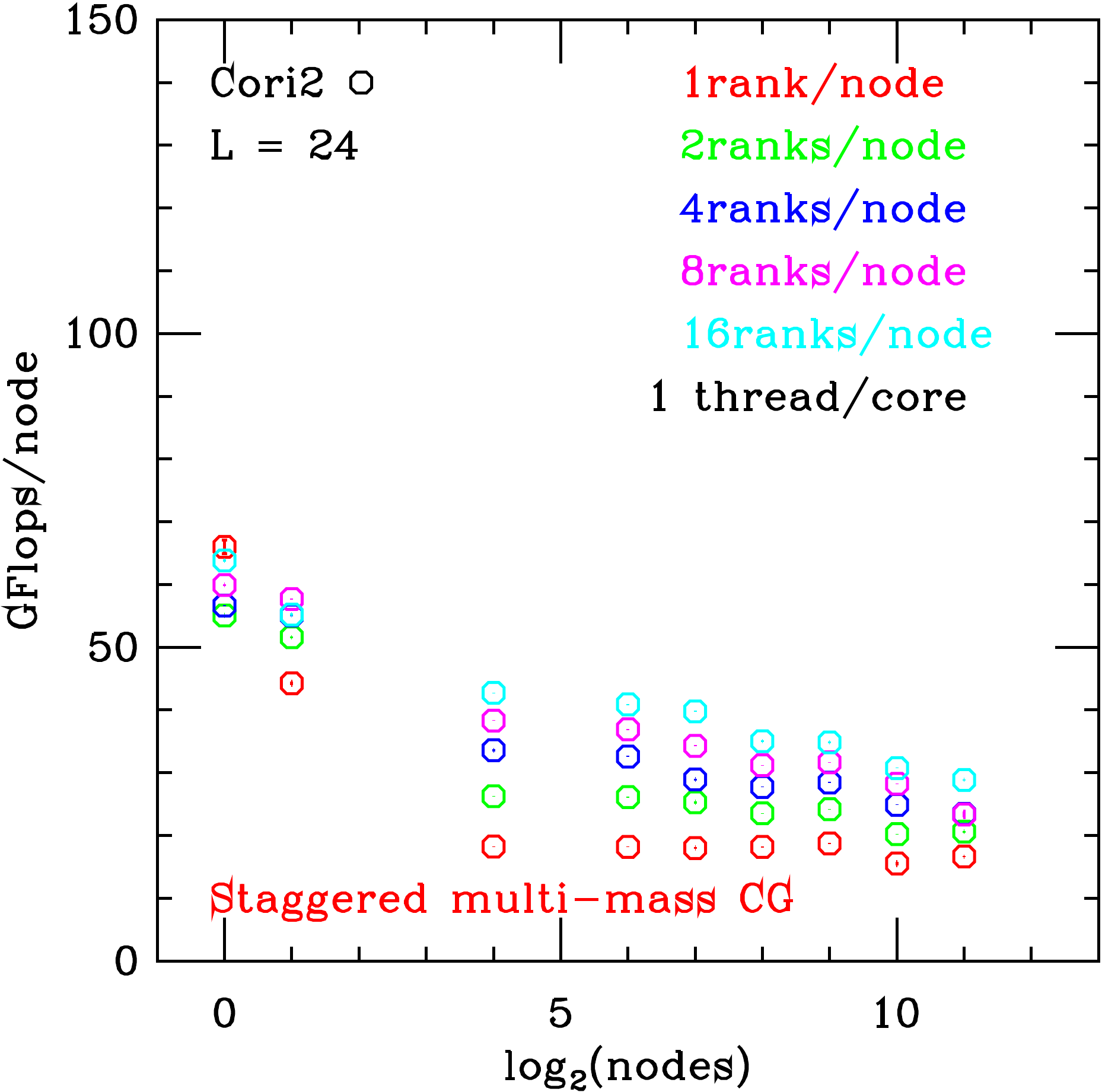}} \hspace{1.1cm} 
   \subfigure[Code with QPhiX performance on multiple nodes does not vary
that much when varying the number of MPI ranks per node. 
             ]%
             {\includegraphics[width=0.32\textwidth,clip]{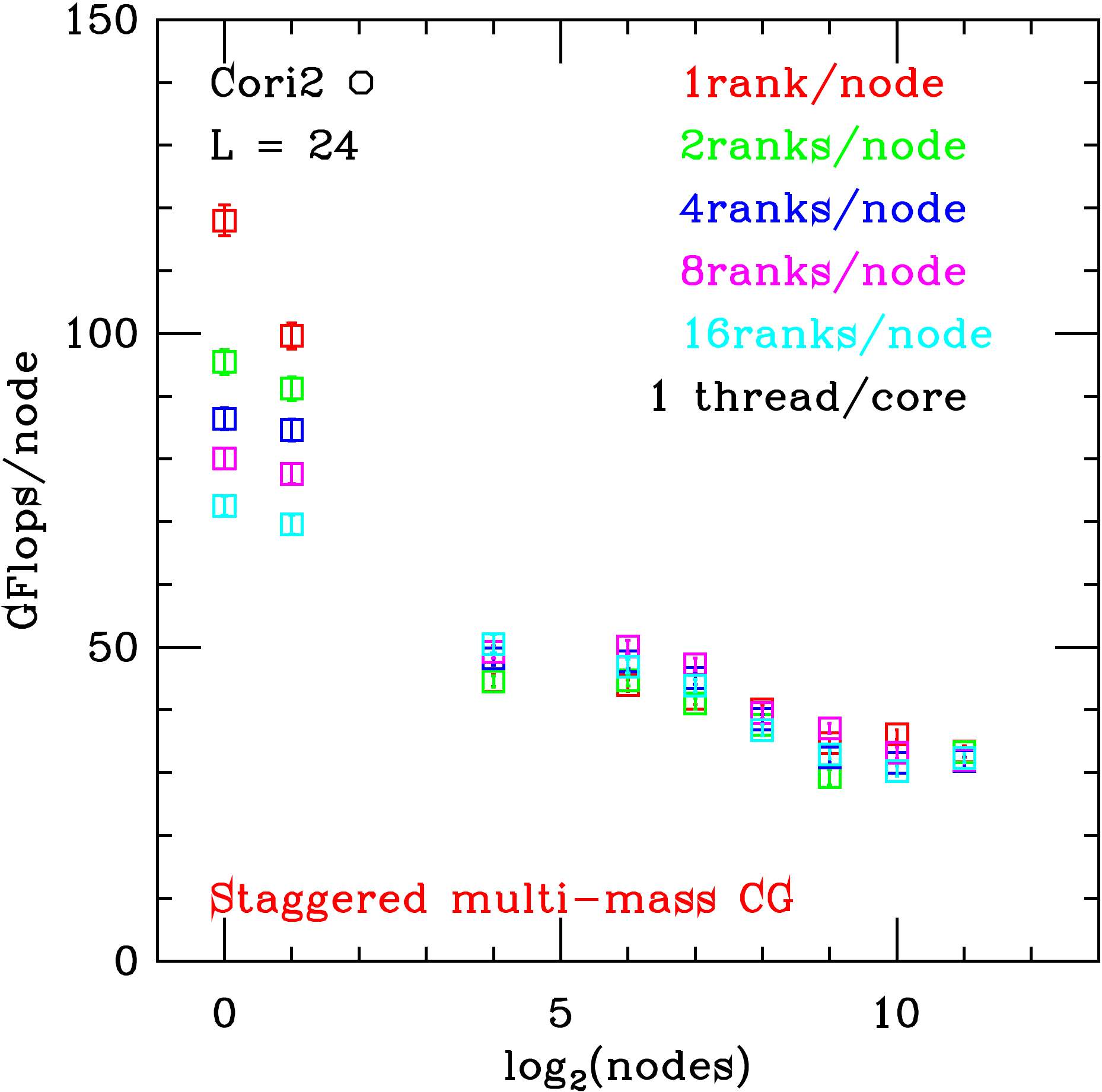}}

   \caption{Staggered multi-shift CG weak-scaling performance on the Cori~II 
KNL cluster using up to 2048 nodes, with baseline code (left) and QPhiX code (right). 
The vertical axis is the flop rate per node. The lattice volume is $24^4$ per node. 
   Runs use no hyper-threading (\emph{i.e.}, 64 total threads per node), and various MPI/OpenMP combinations of up to 16 ranks per node. 
   }
   \label{Fig.3}
\end{figure}

\begin{figure}[thb]
  \centering
  \sidecaption
  \includegraphics[width=0.34\textwidth,clip]{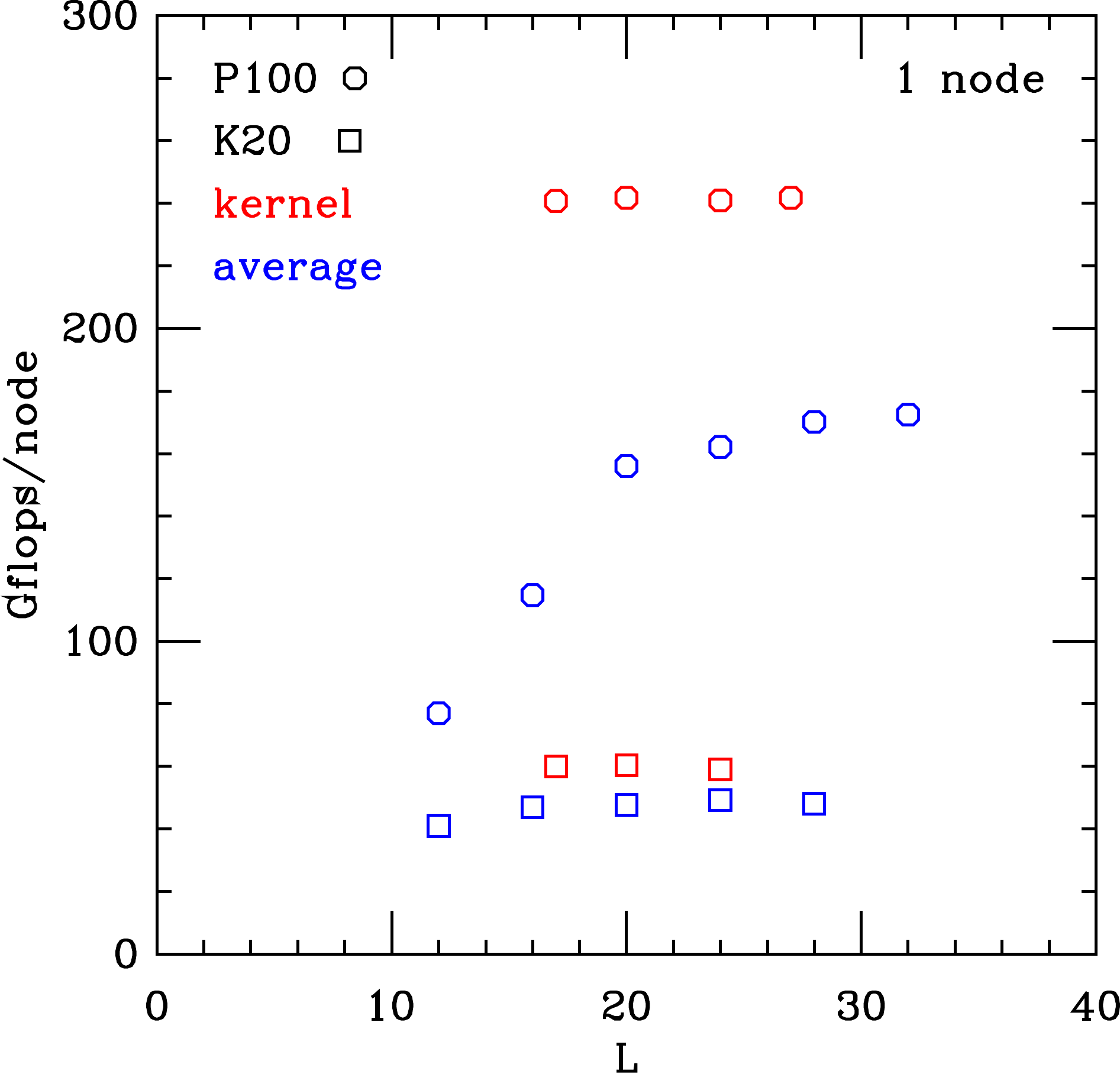}
  \caption{Staggered multi-shift CG performance on one GPU node for various lattice volumes $L^4$. 
  The performance excludes routine overhead, e.g., the data reconstruction and transfer between the host and GPU device. }
  \label{Fig.9}
\end{figure}

\noindent{}Figure \ref{Fig.2} shows the staggered multi-shift CG performance on up to 64 KNL nodes on various clusters, 
comparing baseline code and optimization with QPhiX, 
as well as various lattice volumes $L^4$ per node.
The overall performance improvement with QPhiX was over $50\%$ on a single KNL node. 
We observed up to 120 Gflops/sec. 
On multiple nodes the performance gain with QPhiX was reduced, 
due to a network bottleneck for this routine, 
especially with $L$ less than $30$. \\
\\
Another weak-scaling plot from the Cori~II cluster is shown 
in figure~\ref{Fig.3}. 
This exhibits the same pattern of the performance bottleneck for $L = 24$ 
due to network limitations and the need for cross-node communication. 
The issue is not as severe for a larger lattice size  $L = 32$. 
We observed similar performance with hyper-threading of two threads per core, 
while increasing to four hyper-threads per core negatively impacted the performance. \\
\\
Figure \ref{Fig.9} shows the performance of this routine on one GPU node using
two different NVIDIA architectures, either K20 or P100. 
The kernel QUDA CG performance on one P100 is around 240 Gflops/sec, about five times of that of on single K20. 

\subsection{Symanzik one-loop gauge force and HISQ fermion force}\label{sec-2}

\begin{figure}[tp]
   \centering
   \subfigure[Baseline code on two KNL clusters, with 64 ranks per node. 
             ]%
             {\includegraphics[width=0.32\textwidth,clip]{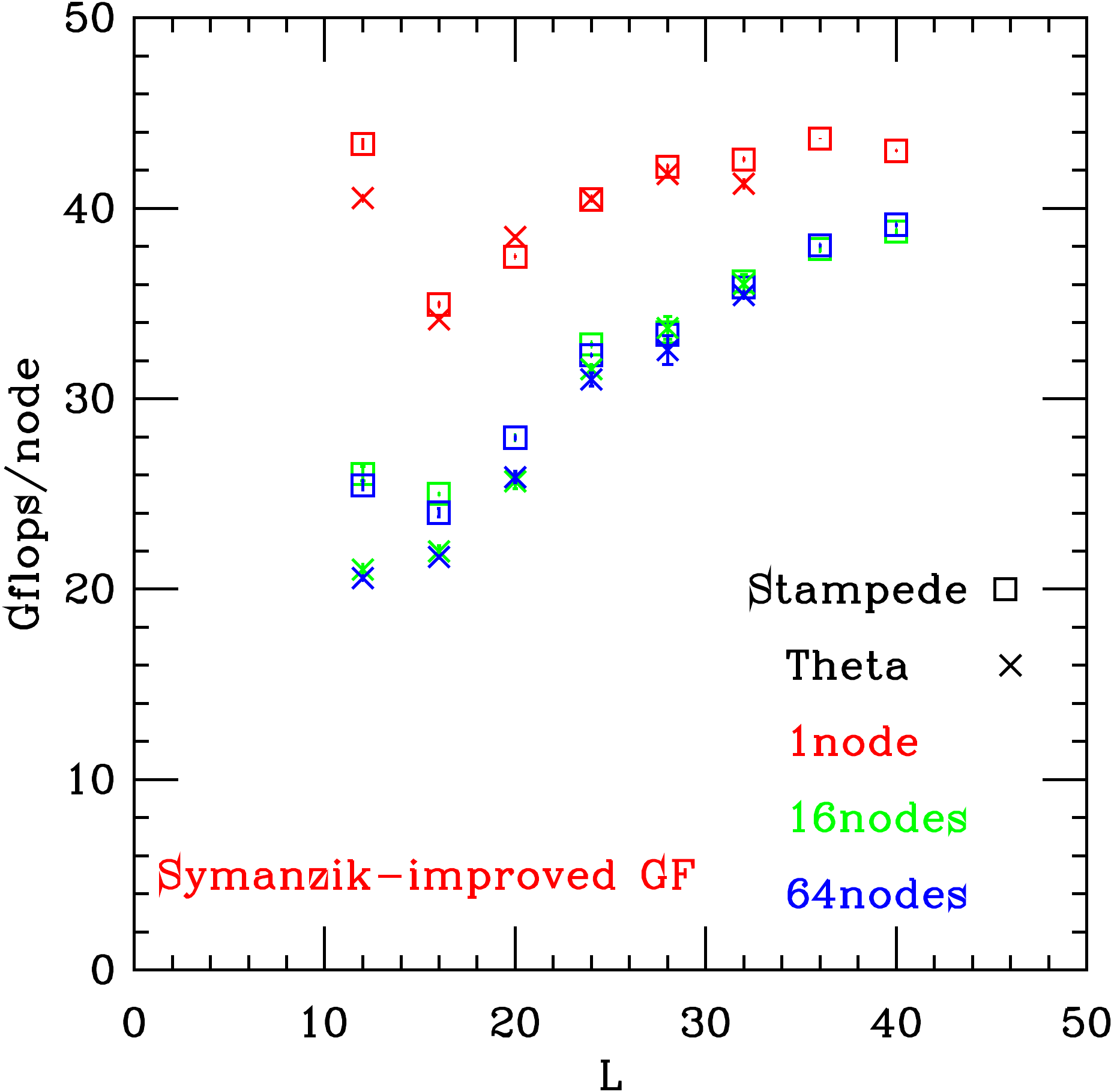}} \hspace{0.3cm} 
   \subfigure[Code with QOPQDP on two KNL clusters, with 64 ranks per node.
   	    ]
   	    {\includegraphics[width=0.30\textwidth,clip]{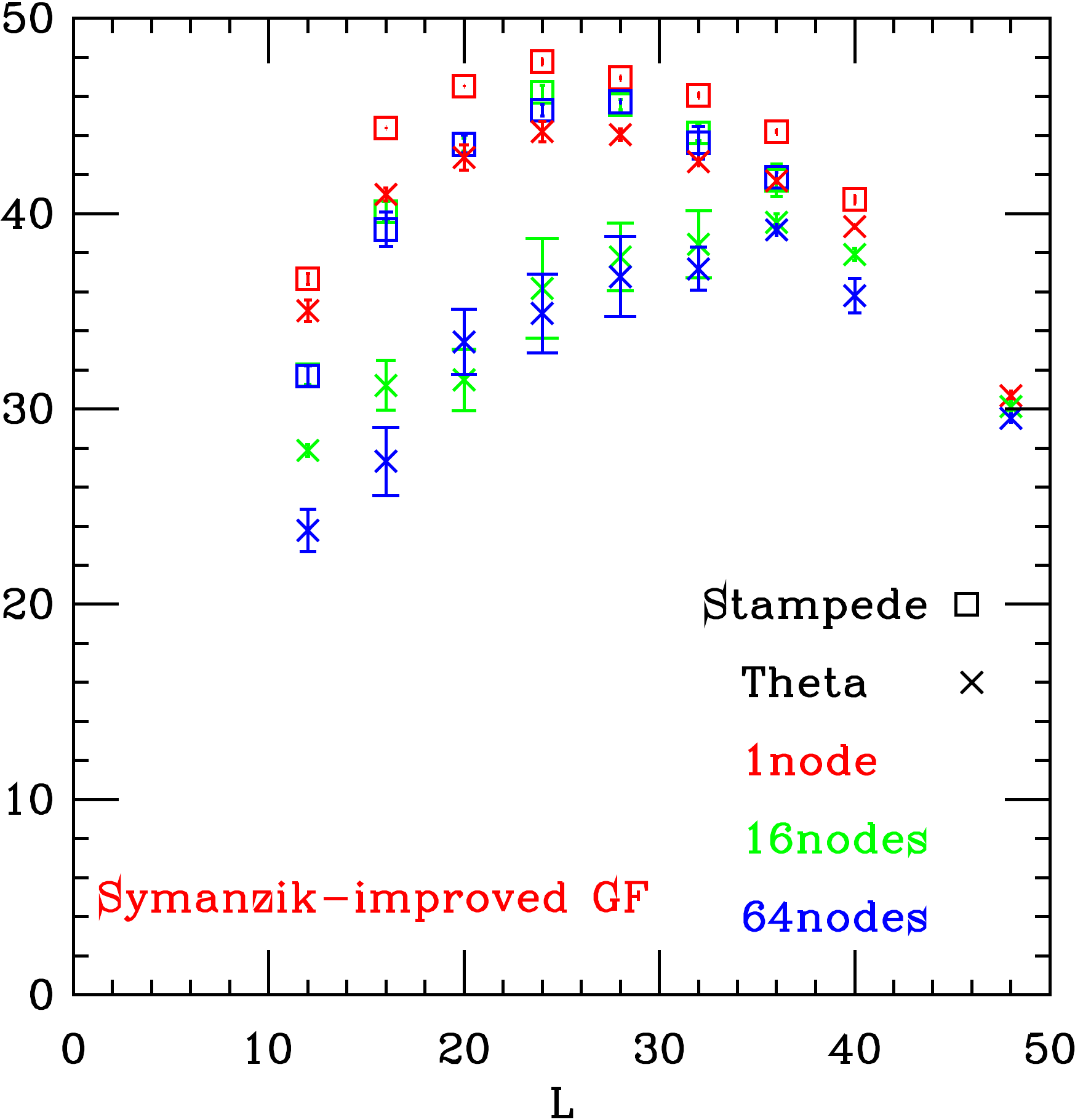}} \hspace{0.3cm} 
   \subfigure[Code with QPhiX on Cori~II cluster.  
   	    One node runs are with OpenMP and 64 threads, 
  	    and multiple-node runs are with 16 ranks per node. 
   	    ]%
            {\includegraphics[width=0.31\textwidth,clip]{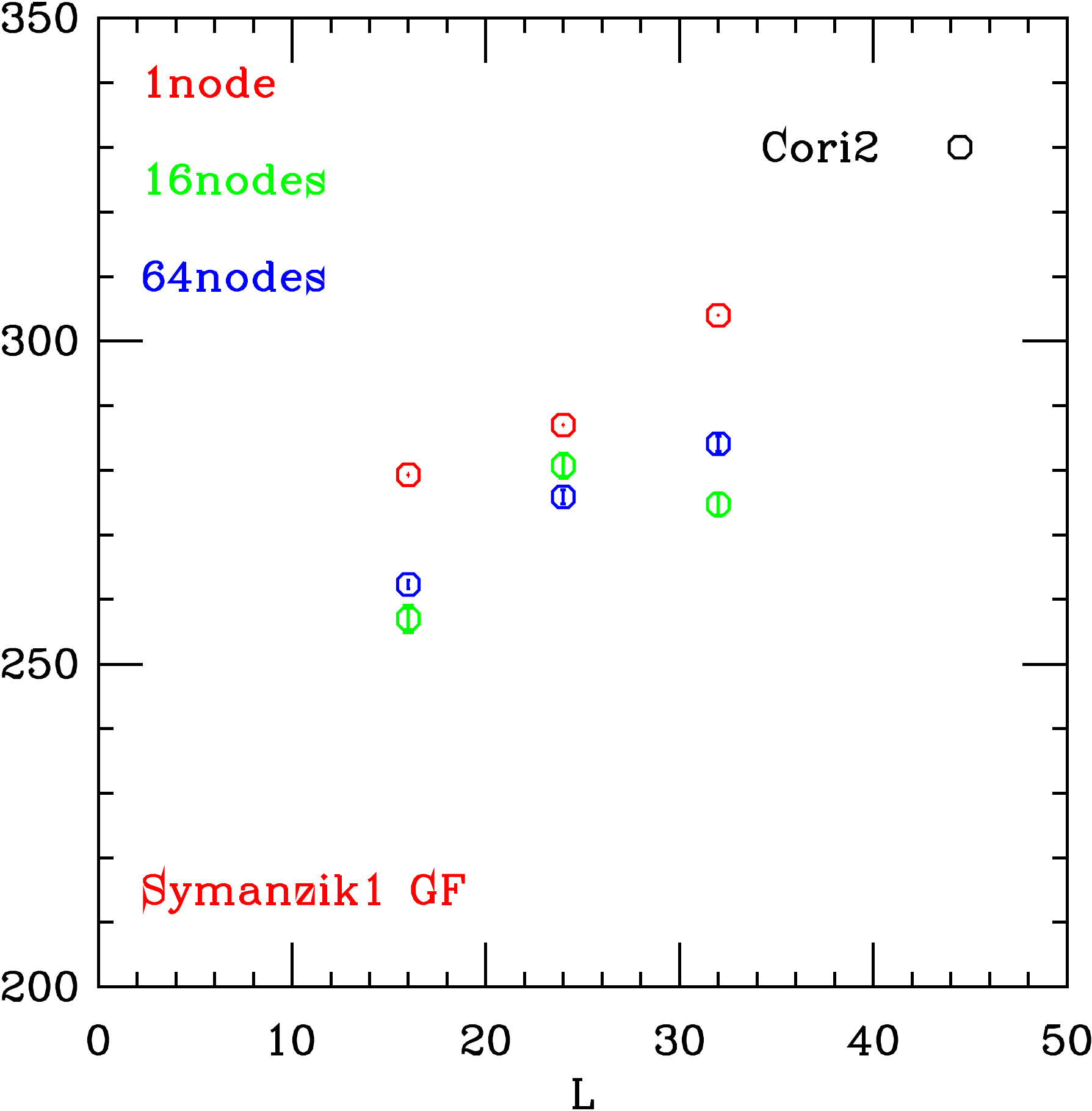}} 
   \caption{Weak scaling of Symanzik one-loop gauge force performance on KNL clusters up to 64 nodes, with baseline code (left), QOPQDP (center), and QPhiX (right). 
   The horizontal axis is the lattice size per node, and vertical axis is the flop rate per node. 
   Each run is with one thread per core. 
   }
   \label{Fig.6}
\end{figure}

The HISQ RHMC/RHMD algorithm also requires the calculation of the gauge force and the fermion force. 
These calculations are dominated by matrix-matrix multiplies,  rather than
by  matrix-vector multiplies as in the CG. 
Thus, they have a higher arithmetic intensity and are less memory 
bandwidth bound than the CG. \\
\\
\noindent{}Figure \ref{Fig.6} shows the gauge-force weak-scaling performance on up to 64 KNL nodes, {\emph vs}.~lattice size $L$. 
The QOPQDP code was compiled to use SSE2 instructions.
Baseline MILC and QOPQDP have similar performance; however, there is 
over five times improvement with QPhiX. 
Using Intel VTune Amplifier to analyze the performance,
we observed a higher cache reuse in the QPhiX algorithm. 
The QOPQDP algorithm also requires fewer flops than baseline MILC by virtue of 
reusing three-link staples.
Its time was reduced to around one third of the baseline time. 
QPhiX also reuses three-link staples and its flop count is reduce by
$60\%$ compared to baseline MILC.  
Thus, the QPhiX timing in total was reduced to less than $6\%$ of that of baseline
MILC, excluding time for data remapping. 
While data remapping currently takes approximately the same amount 
of time as the routine itself, 
we expect to reduce its impact on the HMC performance after all parts of
the algorithm are included in the library, as remapping will not be 
needed every time the routine is called. \\

\begin{figure}[tp]
   \centering
   \subfigure[Baseline code 
   	    ]
   	    {\includegraphics[width=0.315\textwidth,clip]{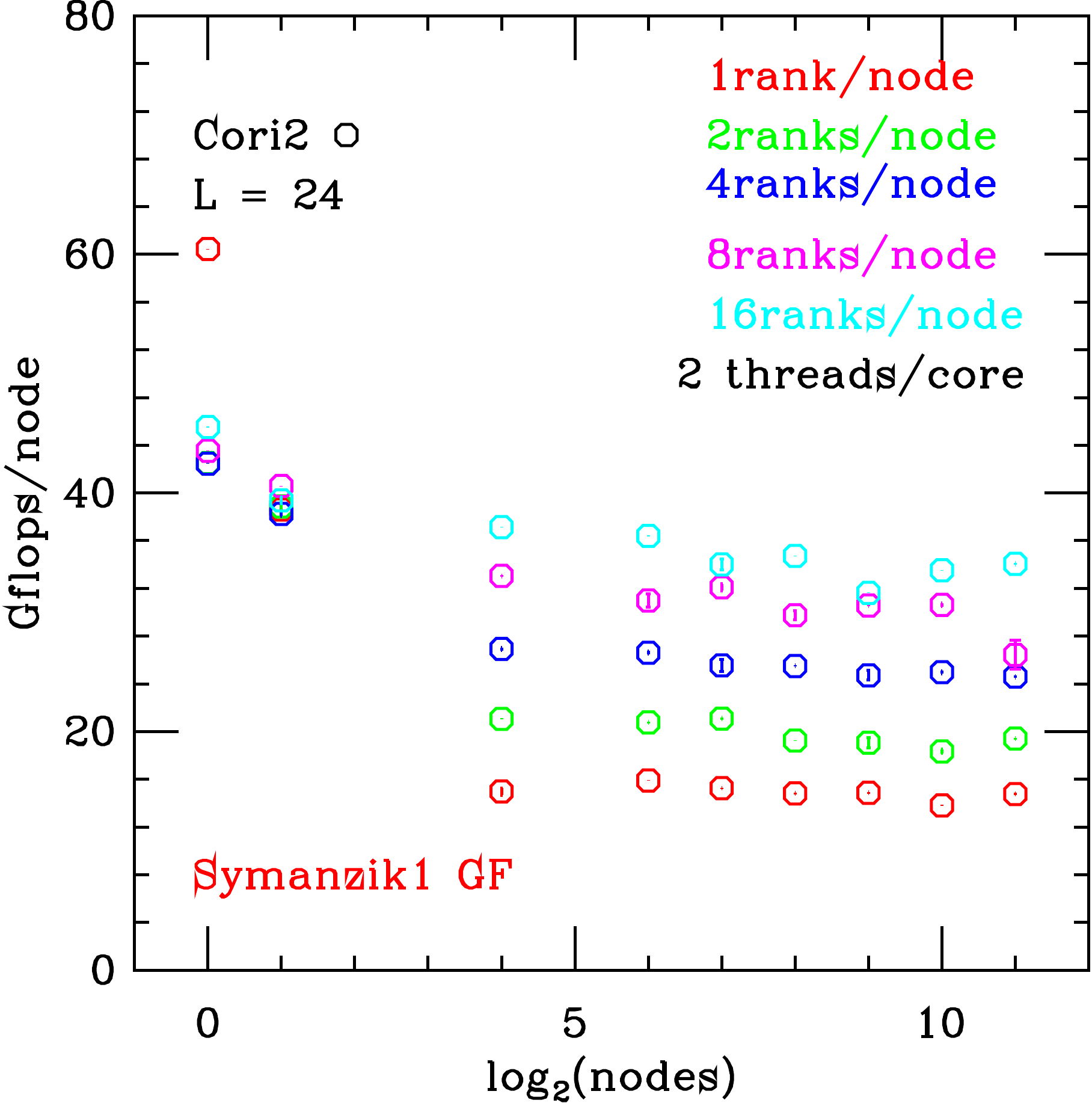}} \hspace{1cm} 
   \subfigure[Code with QPhiX
   	    ]%
            {\includegraphics[width=0.32\textwidth,clip]{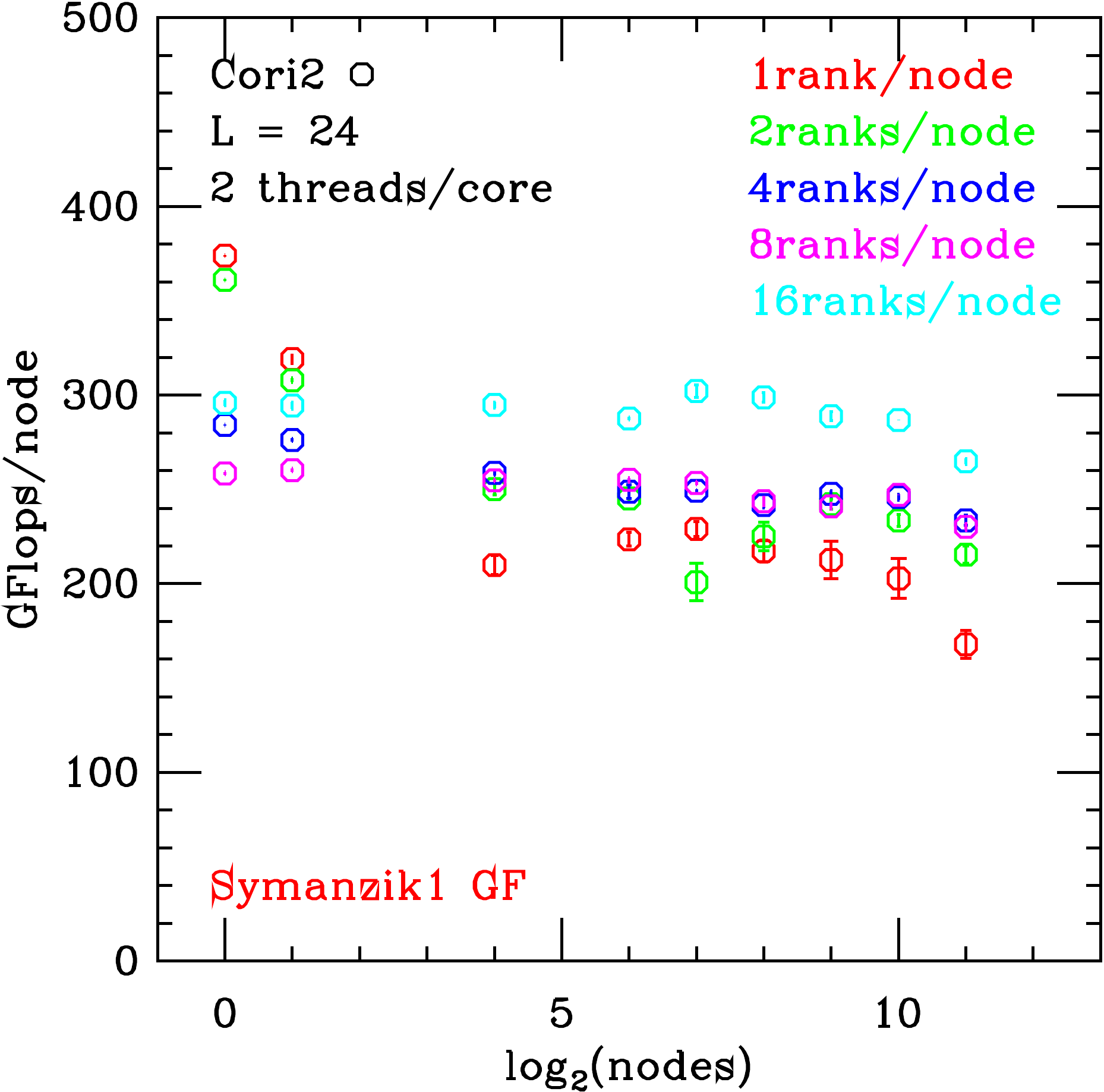}} 
   \caption{Weak scaling study of Symanzik one-loop gauge force performance on the Cori~II cluster with up to 2048 nodes, showing baseline code (left) and QPhiX (right). 
   The vertical axis is the flop rate per node. The lattice size is $24^4$ per node. 
   The runs use two threads per core, and various MPI/OpenMP combinations of up to 16 ranks per node.
   }
   \label{Fig.7}
\end{figure}

\noindent{}Weak scaling of the gauge force
on Cori~II is shown in figure \ref{Fig.7}. 
The scaling efficiency is higher compared with CG: 
over $40\%$ with baseline code and more with QPhiX. 
Also, hyper-threading with two threads per core helps improve the performance. 
The QPhiX improvement in performance here is dramatic compared with that of the CG solver. 
As mentioned, this happens in part because the arithmetic intensity of the gauge-force algorithm is 1.1 in the baseline code 
and over 2.0 in the QPhiX algorithm, 
whereas with the CG, 
it is 0.26 for both baseline and QPhiX codes.  
We also rerranged the algorithm
to avoid frequent communication,
thus leading to a higher weak scaling efficiency.
On the other hand, 
the performance of QPhiX at different MPI/OpenMP combinations varies more than the CG performance does. 

\begin{figure}[thb]
  \centering
  \sidecaption
  \includegraphics[width=0.34\textwidth,clip]{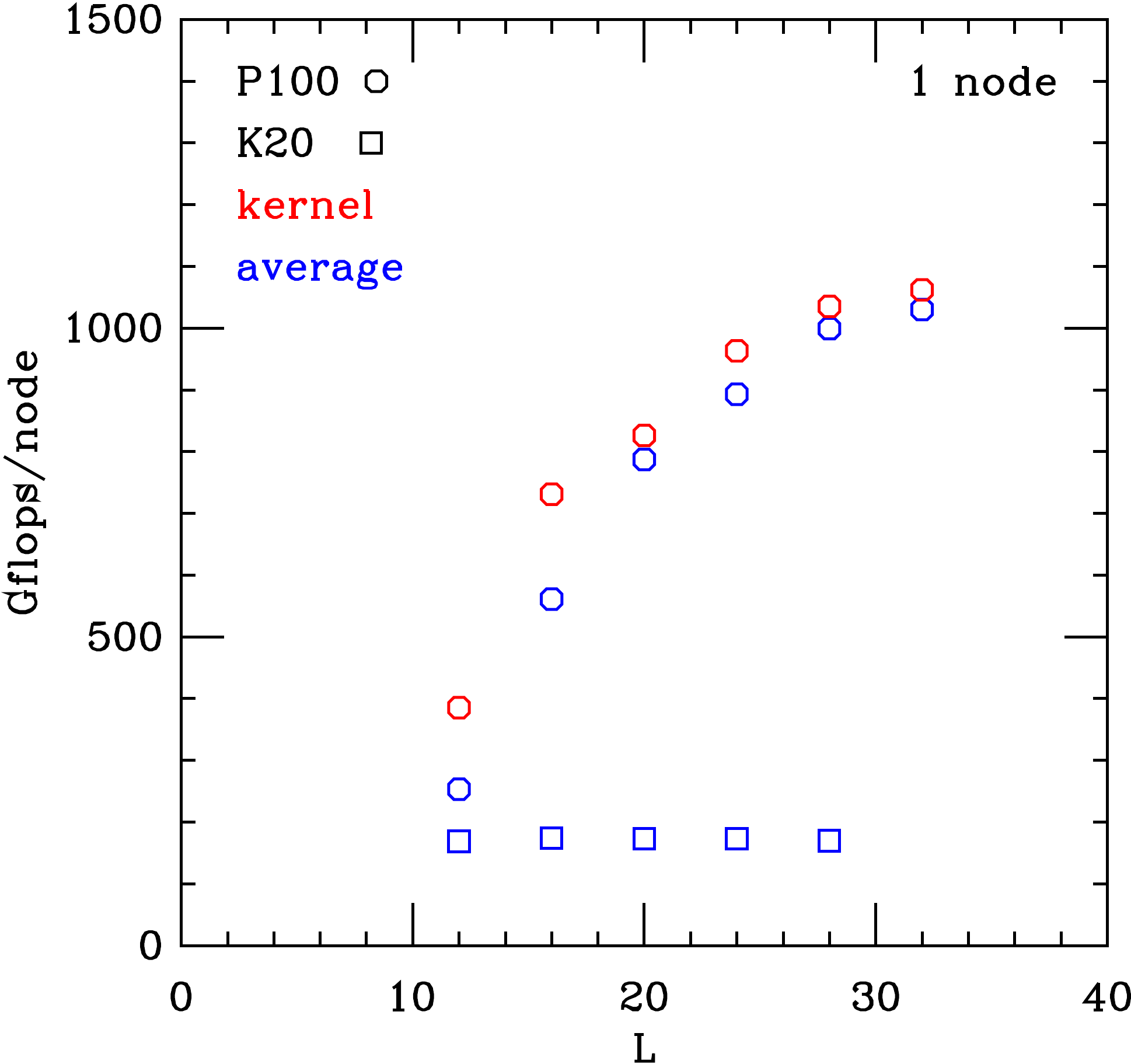}
  \caption{Symanzik one-loop gauge force performance on one GPU node and various lattice sizes $L$. 
  The performance excludes data remapping overhead.}
  \label{Fig.8}
\end{figure}

\noindent{}The gauge force performance on one GPU node is shown in figure \ref{Fig.8}, 
on various lattice sizes. 
Comparing P100 and K20, the performance ratio on these two architectures is again up to 5. 
On the other hand, the performance on one P100 is around $2.3$ times that on one KNL, 
which is slightly higher than the ratio of CG on these two architectures. 
The algorithm in QUDA is the same as in the baseline code. \\


\noindent{}The HISQ fermion force routine was also benchmarked, 
comparing the performance of the baseline code and the QOPQDP library. 
We observed similar flop rates again for these two versions of the code, 
for which single node runs were around 40--50 Gflops/sec with the 
baseline code and 30--40 Gflops/sec with QOPQDP. 
The weak-scaling performance efficiency is 50--80\% on up to 64 KNL nodes. 
Because of the reduced amount of computation in the QOPQDP routine, 
its timing was reduced to around $20\%$ of that in the baseline code.

\section{Conclusions}\label{sec-1}

We explored performance of three major lattice QCD routines in the MILC code: 
the staggered multi-shift CG, Symanzik one-loop gauge-force, and the HISQ fermion force. 
We ran our benchmarks on KNL clusters located at three scientific supercomputer centers: ALCF, NERSC, and TACC. 
We compared the staggered multishift CG and Symanzik gauge force performance from optimizations with QPhiX, 
with that of the baseline MILC code and the QOPQDP library. 
We found the CG performance with QPhiX was improved by $1.5$ times on one KNL, 
while the algorithm was network bandwidth bound on clusters. 
On the other hand, 
RHMD routines such as the gauge force and fermion force showed a weak scaling 
efficiency of up to over $80\%$. 
Comparing the performance improvement of these routines, 
we argued a higher arithmetic intensity of the gauge force calculations in QPhiX contributed to a higher performance improvement (over ten times) of this routine. 
As a comparison, we also showed the code performance on single GPU K20 and P100 nodes with the QUDA 0.8.0 library. 
In general, the optimized routine performance of a single P100 is 
about five times of that of a K20. 
We are developing QPhiX versions of other parts of the MILC code,
including the HISQ fermion force and the HISQ link smearing, to add to the library.

\clearpage

\end{document}